\documentclass[runningheads]{llncs}
\usepackage{ecir25-densereviewer-demo-frame}
\graphicspath{{Fig/}}

\begin{document}
\title{DenseReviewer: A Screening Prioritisation Tool for Systematic Review based on Dense Retrieval}
\titlerunning{DenseReviewer: Dense Retrieval and Relevance Feedback Screening}

\author{
Xinyu Mao\inst{1}\orcidlink{0000-0001-6357-2311} \and
Teerapong Leelanupab\inst{1}\orcidlink{0000-0002-8117-0612} \and\\
Harrisen Scells\inst{2}\orcidlink{0000-0001-9578-7157} \and 
Guido Zuccon\inst{1}\orcidlink{0000-0003-0271-5563}}
\authorrunning{Mao et al.}
	
\institute{The University of Queensland \and University of Kassel and hessian.AI}

\maketitle

\begin{abstract}
\sloppy
Screening is a time-consuming and labour-intensive yet required task for medical systematic reviews, as tens of thousands of studies often need to be screened. Prioritising relevant studies to be screened allows downstream systematic review creation tasks to start earlier and save time. In previous work, we developed a dense retrieval method to prioritise relevant studies with reviewer feedback during the title and abstract screening stage. Our method outperforms previous active learning methods in both effectiveness and efficiency. 
In this demo, we extend this prior work by creating (1) a web-based screening tool that enables end-users to screen studies exploiting state-of-the-art methods and (2) a Python library that integrates models and feedback mechanisms and allows researchers to develop and demonstrate new active learning methods.
We describe the tool's design and showcase how it can aid screening. The tool is available at \url{https://densereviewer.ielab.io}. The source code is also open sourced at \url{https://github.com/ielab/densereviewer}.

\end{abstract}
\setcounter{footnote}{0} 
\section{Introduction and Related Work}
Medical systematic reviews (SRs) synthesise evidence from the literature, requiring high recall to avoid missing relevant studies. The screening process is critical to ensure high recall and is a two-stage process: Firstly, the title and abstract of studies are assessed by medical researchers or librarians for relevance, followed by the full text. The former title and abstract (T\&A) screening generally involves tens of thousands of studies~\cite{borah2017analysis}, leading to a high workload and cost. Several tools and products have been developed to reduce this workload, including
ASReview~\cite{van2021open},%
\footnote{\url{https://asreview.nl/}} 
Covidence,%
\footnote{\url{https://www.covidence.org/}}
DistillerSR,%
\footnote{\url{https://www.distillersr.com/}}
and RobotAnalyst~\cite{przybyla2018prioritising}.%
\footnote{\url{https://nactem.ac.uk/robotanalyst/}}
These tools classify studies using classical machine learning. Each study is suggested for inclusion or exclusion or labelled with a confidence score by the model. 

Prior work has proposed to use active learning (AL)~\cite{settles2009active} to strategically select studies for manual judgement in order to iteratively train models to more effectively prioritise relevant studies. The use of AL in systematic review automation tools is so far limited~\cite{van2021open}. Furthermore, recent studies~\cite{yang2022goldilocks,mao2024reproducibility} showed that neural models such as BERT have the potential to prioritise studies much more effectively than previous approaches using AL, especially when pre-trained on domain-specific data (e.g., bio-medicine)~\cite{mao2024reproducibility}. However, one downside to these highly effective models, and in fact all AL methods, is that they still require bootstrapping in the form of pre-selected relevant studies. Their computational cost also makes them considerably slower than traditional classification methods. 

In this paper, we demonstrate DenseReviewer, a screening tool leveraging dense retrievers and tailored queries (i.e., PICO: patient/population, intervention, comparison, and outcome~\cite{scells2017integrating}) for T\&A screening. Key features of DenseReviewer are summarised in table~\ref{tab:tool-compare}, and compared with popular SR tools mentioned above.
DenseReviewer iteratively updates a PICO query efficiently via the Rocchio's algorithm for dense retrieval~\cite{li2023pseudo} based on the screener's feedback (i.e., the judgement of each screened study). 
Our previous work~\cite{mao2024dense} showed that this dense retrieval-based approach is more effective for screening than logistic regression-based and BERT-based active learning workflows, while maintaining efficiency comparable to traditional machine learning-based methods. 

\begin{table}[t!]
	\centering
	\caption{Comparison of key features between DenseReviewer and popular SR tools. `Full Screen' shows the title and abstract of one study at a time, while `Ranking List' presents studies with their titles and abstracts for screening in an order, typically by relevance.}
	\begin{tabular}{p{9em}p{13em}p{8em}ll@{}}
		\toprule
		\textbf{Screening Tool} & \textbf{Core Technology} & \textbf{Interface} & \textbf{Code} \\ \midrule
		ASReview       & \multirow{4}{*}{\shortstack[l]{Machine Learning \\ and Active Learning \\ (model training)}}         & Full Screen          & Open Source   \\
		Covidence     &         & Ranking List        & Proprietary   \\
		DistillerSR      &       & Ranking List        & Proprietary   \\
		RobotAnalyst  &     & Ranking List         & Proprietary   \\ \midrule
		\raisebox{+0.5\totalheight}{DenseReviewer}   & {\shortstack[l]{Dense Retrieval \\ and Relevance Feedback \\ (query vector updating)}}         & {\shortstack[l]{Ranking List \\ and Full Screen}}     & \raisebox{+0.5\totalheight}{Open Source}   \\ \bottomrule
	\end{tabular}
	\label{tab:tool-compare}
\end{table}

\begin{figure*}[t!]
\centering
\includegraphics[width=\columnwidth]{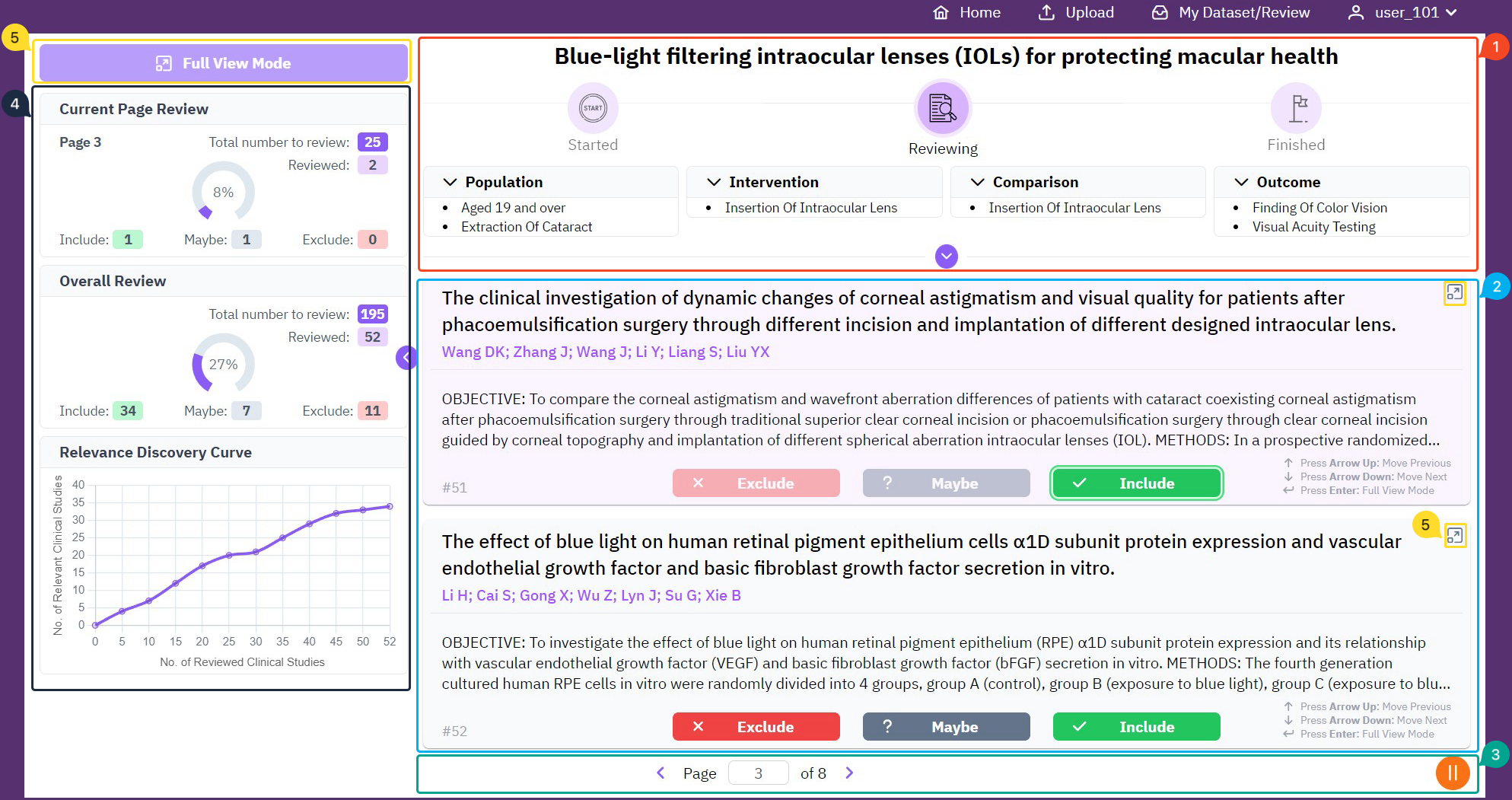}
\caption{Ranking mode. \circled{1}{Red} contains the PICO query. \circled{2}{Blue} lists the studies; users can use keyboard or mouse controls to expand a study to read and judge. Assessed studies are highlighted in purple. \circled{3}{Green} contains page controls, with a pause button to save the review's progress and toggles to stop upon reaching the last page. \circled{4}{Navy} shows two pie charts that display the ratio of reviewed to unreviewed studies and the distribution of judgements and a line chart that displays the relevance discovery curve, indicating the saturation of relevant studies throughout the screening progresses. \circled{5}{Yellow} allows users to enter focus mode (see Figure~\ref{fig:ui_full_screen_mode}).}
\label{fig:ui_ranking_mode}
\end{figure*}

\begin{figure*}[t!]
\centering
\includegraphics[width=\textwidth]{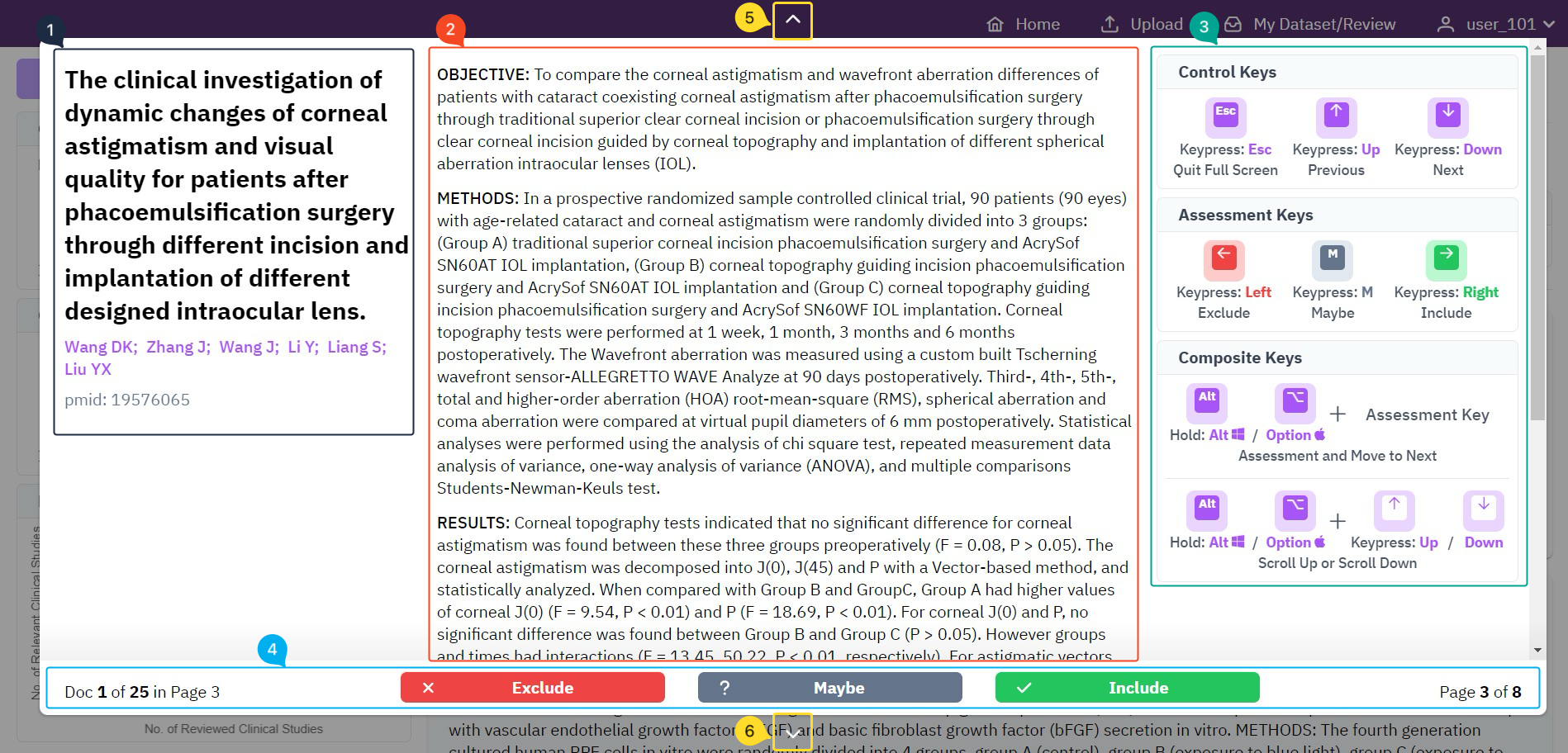}
\caption{Focus mode. \circled{1}{Navy} displays the study's title, list of authors, and PubMed ID. \circled{2}{Red}~contains the full abstract. \circled{4}{Blue} includes the three assessment options, a ranking index of the current study on the page, and the page number, positioned centrally, to the left, and right, respectively. Buttons \circled{5}{Yellow} and \circled{6}{Yellow} allow users to navigate between studies.}
\label{fig:ui_full_screen_mode}
\end{figure*}

\section{Overview of DenseReviewer}
DenseReviewer offers two modes: ranking mode and focus mode.
Figure~\ref{fig:ui_ranking_mode} provides an overview of ranking mode, allowing users to browse and make assessments on studies from the perspective of a paginated, ranked list. Figure~\ref{fig:ui_full_screen_mode} provides an overview of focus mode, allowing users to review each study individually and efficiently. In this mode, keyboard controls can be used to quickly move between studies and make assessments. 

Users can upload their corpus, retrieved from PubMed in \texttt{nbib} format, and submit a structured PICO query~\cite{scells2017integrating}. Uploaded studies must include the pmid (PubMed identifier), title, abstract and a list of authors. A dense retriever ranks uploaded studies with respect to the PICO query. After assessing each page of studies, the remaining unjudged studies are re-ranked based on Rocchio's algorithm with positive (include) and negative (exclude) feedback~\cite{mao2024dense}. DenseReviewer is open source, and all its components, including the front-end, API back-end, and database, are packaged within Docker containers. The containers can be downloaded from a single git repository and self-hosted with ease. Aside from the GUI-based system for screening, we provide a Python library%
\footnote{\url{https://github.com/ielab/dense-screening-feedback}} 
to enable experimentation using existing datasets~\cite{wang2022little,scells2017test,kanoulas2017clef,kanoulas2018clef,kanoulas2019clef}, and training or loading dense retrievers with customised backbone models.
\section{Architecture and Library}
DenseReviewer comprises six Docker containers that can all be deployed on a single cloud instance:\enlargethispage{2\baselineskip} (i) a web-based front end, (ii) a REST API back end, (iii) a database for storing information such as user activity, uploaded corpora, and ranking studies, (iv) a message broker for managing asynchronous task queues, (v) a service dedicated to tasks such as parsing, encoding, indexing, and initial ranking when new datasets are uploaded to review, and (vi) a service responsible for handling re-ranking. We deployed DenseReviewer on an AWS EC2 accelerated computing instance (g4dn.xlarge: 1 T4 Tensor GPU, 16 GB GPU memory, 4 vCPUs, and 16 GB instance memory). The architecture can scale to more powerful instances, such as the G5 or P4 series with NVIDIA A10G and A100 GPUs, without requiring major modifications. 

\begin{figure}[t!]
\lstset{basicstyle=\ttfamily, columns=fullflexible, keepspaces=true,frame=tb}
\scriptsize
\newsavebox{\training}
\begin{lrbox}{\training}
\begin{minipage}{.48\textwidth}
\begin{lstlisting}
python tevatron_pipe.py \
	--collection_split clef19_dta_train \
	--model_path biolinkbert \
	--q_max_len 128 \
	--p_max_len 256 \
	--train_n 11 \
	--train_epoch 60
	
	
\end{lstlisting}
\end{minipage}
\end{lrbox}
\newsavebox{\screening}
\begin{lrbox}{\screening}
\begin{minipage}{.48\textwidth}
\begin{lstlisting}
python dense_query_tar.py \
	--collection_split clef19_dta_test \
	--model biolinkbert_128_256_11 \
	--n_iteration 20 \
	--top_k 20 \
	--output_path ourput_dir \
	--alpha 1.0 \
	--beta 0.8 \
	--gamma 0.2
\end{lstlisting}
\end{minipage}
\end{lrbox}
\subcaptionbox{Training\label{fig:python.training}}[0.48\textwidth]{\usebox{\training}}
\subcaptionbox{Screening\label{fig:python.screening}}[0.48\textwidth]{\usebox{\screening}}
\setcounter{figure}{2}    
\captionof{figure}{Command line usage of DenseReviewer for training and screening.}
\label{fig:python}
\end{figure}

Figure~\ref{fig:python} shows the command line usage for (i) training dense retrievers based on \texttt{Tevatron} (Figure~\ref{fig:python.training}) and (ii)  running experiments with the trained dense retrievers and feedback methods (Figure~\ref{fig:python.screening}).

\section{Conclusion and Future Work}
\label{future}
\sloppy
In this demonstration paper, we introduced DenseReviewer, a framework for title and abstract screening for medical systematic reviews with dense retrieval and active learning. We presented a web application, which showcased two interfaces to screen studies. We also presented the underlying Python library which showcased dense retrieval active learning experimentation.

Going forward, we have several new functionalities planned. 
The first planned functionality is to extend the PICO query to allow users to express exclusion criteria. 
The second planned functionality is the automatic extraction and highlighting of potential words or sentences associated with (non-)relevant PICO elements. Several studies have shown the potential of large language models (LLMs) in supporting this task~\cite{ghosh2024alpapico,reason2024automated}. 
The third planned functionality is to automate relevance judgements during screening. We will investigate LLM-assisted screening to further reduce screeners' workload. Several studies have shown the applicability of LLMs at this task~\cite{cao2024prompting,bron2024combining,scherbakov2024emergence}.


\section*{Acknowledgments}
\label{ack}
Xinyu Mao is supported by a UQ Earmarked PhD Scholarship and this research is funded by the Australian Research Council Discovery Projects programme ARC DP DP210104043.
We extend our gratitude to the engineering team of AI~DETA Technologies~Co.\footnote{\url{https://aideta.com/}} (previously named Thaibiogenix~Co.), including Kanlayakorn Yeenang, Weeravat Buachoom, Tasanai Srisawat, Warangkhana Sukpartcharoen, and Thuchpun Apivitcholachat for their consultation and support in developing the web application for deploying  the \texttt{DenseReviewer} core.
\bibliographystyle{splncs04}
\bibliography{Bib/bibliography}

\begin{thebibliography}{10}
\providecommand{\url}[1]{\texttt{#1}}
\providecommand{\urlprefix}{URL }
\providecommand{\doi}[1]{https://doi.org/#1}

\bibitem{borah2017analysis}
Borah, R., Brown, A.W., Capers, P.L., Kaiser, K.A.: Analysis of the time and
  workers needed to conduct systematic reviews of medical interventions using
  data from the prospero registry. BMJ open  \textbf{7}(2) (2017)

\bibitem{bron2024combining}
Bron, M.P., Greijn, B., Coimbra, B.M., van~de Schoot, R., Bagheri, A.:
  Combining large language model classifications and active learning for
  improved technology-assisted review  (2024)

\bibitem{cao2024prompting}
Cao, C., Sang, J., Arora, R., Kloosterman, R., Cecere, M., Gorla, J., Saleh,
  R., Chen, D., Drennan, I., Teja, B., et~al.: Prompting is all you need: Llms
  for systematic review screening. medRxiv pp. 2024--06 (2024)

\bibitem{ghosh2024alpapico}
Ghosh, M., Mukherjee, S., Ganguly, A., Basuchowdhuri, P., Naskar, S.K.,
  Ganguly, D.: Alpapico: Extraction of pico frames from clinical trial
  documents using llms. Methods  \textbf{226},  78--88 (2024)

\bibitem{kanoulas2017clef}
Kanoulas, E., Li, D., Azzopardi, L., Spijker, R.: Clef 2017 technologically
  assisted reviews in empirical medicine overview. In: CEUR Workshop
  Proceedings. vol.~1866, pp. 1--29 (2017)

\bibitem{kanoulas2018clef}
Kanoulas, E., Li, D., Azzopardi, L., Spijker, R.: Clef 2018 technologically
  assisted reviews in empirical medicine overview. In: CEUR Workshop
  Proceedings. vol.~2125 (2018)

\bibitem{kanoulas2019clef}
Kanoulas, E., Li, D., Azzopardi, L., Spijker, R.: Clef 2019 technology assisted
  reviews in empirical medicine overview. In: CEUR Workshop Proceedings.
  vol.~2380 (2019)

\bibitem{li2023pseudo}
Li, H., Mourad, A., Zhuang, S., Koopman, B., Zuccon, G.: Pseudo relevance
  feedback with deep language models and dense retrievers: Successes and
  pitfalls. ACM Transactions on Information Systems  \textbf{41}(3),  1--40
  (2023)

\bibitem{mao2024reproducibility}
Mao, X., Koopman, B., Zuccon, G.: A reproducibility study of goldilocks:
  Just-right tuning of bert for tar. In: European Conference on Information
  Retrieval. pp. 132--146. Springer (2024)

\bibitem{mao2024dense}
Mao, X., Zhuang, S., Koopman, B., Zuccon, G.: Dense retrieval with continuous
  explicit feedback for systematic review screening prioritisation. In:
  Proceedings of the 47th International ACM SIGIR Conference on Research and
  Development in Information Retrieval. pp. 2357--2362 (2024)

\bibitem{przybyla2018prioritising}
Przyby{\l}a, P., Brockmeier, A.J., Kontonatsios, G., Le~Pogam, M.A., McNaught,
  J., von Elm, E., Nolan, K., Ananiadou, S.: Prioritising references for
  systematic reviews with robotanalyst: a user study. Research synthesis
  methods  \textbf{9}(3),  470--488 (2018)

\bibitem{reason2024automated}
Reason, T., Langham, J., Gimblett, A.: Automated mass extraction of over
  680,000 picos from clinical study abstracts using generative ai: A
  proof-of-concept study. Pharmaceutical Medicine pp.~1--8 (2024)

\bibitem{scells2017integrating}
Scells, H., Zuccon, G., Koopman, B., Deacon, A., Azzopardi, L., Geva, S.:
  Integrating the framing of clinical questions via pico into the retrieval of
  medical literature for systematic reviews. In: CIKM'17 (2017)

\bibitem{scells2017test}
Scells, H., Zuccon, G., Koopman, B., Deacon, A., Azzopardi, L., Geva, S.: A
  test collection for evaluating retrieval of studies for inclusion in
  systematic reviews. In: Proceedings of the 40th international ACM SIGIR
  conference on research and development in information retrieval. pp.
  1237--1240 (2017)

\bibitem{scherbakov2024emergence}
Scherbakov, D., Hubig, N., Jansari, V., Bakumenko, A., Lenert, L.A.: The
  emergence of large language models (llm) as a tool in literature reviews: an
  llm automated systematic review. arXiv preprint arXiv:2409.04600  (2024)

\bibitem{settles2009active}
Settles, B.: Active learning literature survey  (2009)

\bibitem{van2021open}
Van De~Schoot, R., De~Bruin, J., Schram, R., Zahedi, P., De~Boer, J., Weijdema,
  F., Kramer, B., Huijts, M., Hoogerwerf, M., Ferdinands, G., et~al.: An open
  source machine learning framework for efficient and transparent systematic
  reviews. Nature machine intelligence  \textbf{3}(2),  125--133 (2021)

\bibitem{wang2022little}
Wang, S., Scells, H., Clark, J., Koopman, B., Zuccon, G.: From little things
  big things grow: A collection with seed studies for medical systematic review
  literature search. In: Proceedings of the 45th International ACM SIGIR
  Conference on Research and Development in Information Retrieval. pp.
  3176--3186 (2022)

\bibitem{yang2022goldilocks}
Yang, E., MacAvaney, S., Lewis, D.D., Frieder, O.: Goldilocks: Just-right
  tuning of bert for technology-assisted review. In: European Conference on
  Information Retrieval. pp. 502--517. Springer (2022)

\end{thebibliography}
\end{document}